\documentclass[twocolumn,prl,superscriptaddress,floatfix,showpacs]{revtex4}
\usepackage{graphicx,bm}
\usepackage[psamsfonts]{amssymb}
\usepackage{type1cm}
\usepackage{dcolumn}
\usepackage{amsmath}

\begin{document}

\title{Evidence of silicene in honeycomb structures of silicon on Ag(111)}

\author{Baojie Feng}
\affiliation{Institute of Physics, Chinese Academy of Sciences,
Beijing 100190, China}
\author{Zijing Ding}
\affiliation{Institute of Physics, Chinese Academy of Sciences,
Beijing 100190, China}
\author{Sheng Meng }
\affiliation{Institute of Physics, Chinese Academy of Sciences,
Beijing 100190, China}
\author{Yugui Yao}
\affiliation{School of Physics, Beijing Institute of
Technology,Beijing 100081, China} \affiliation{Institute of
Physics, Chinese Academy of Sciences, Beijing 100190, China}
\author{Xiaoyue He}
\affiliation{Institute of Physics, Chinese Academy of Sciences,
Beijing 100190, China}
\author{Peng Cheng}
\affiliation{Institute of Physics, Chinese Academy of Sciences,
Beijing 100190, China}
\author{Lan Chen\,\footnote{Corresponding author, Email: lchen@iphy.ac.cn}}
\affiliation{Institute of Physics, Chinese Academy of Sciences,
Beijing 100190, China}
\author{Kehui Wu\,\footnote{Corresponding author, Email: khwu@iphy.ac.cn}}
\affiliation{Institute of Physics, Chinese Academy of Sciences,
Beijing 100190, China}

\date{\today}

\begin{abstract}
In the search for evidence of silicene, a two-dimensional
honeycomb lattice of silicon, it is important to obtain a complete
picture for the evolution of Si structures on Ag(111), which is
believed to be the most suitable substrate for growth of silicene
so far. In this work we report the finding and evolution of
several monolayer superstructures of silicon on Ag(111) depending
on the coverage and temperature. Combined with first-principles
calculations, the detailed structures of these phases have been
illuminated. These structure were found to share common building
blocks of silicon rings, and they evolve from a fragment of
silicene to a complete monolayer silicene and multilayer silicene.
Our results elucidate how silicene formes on Ag(111) surface and
provide methods to synthesize high-quality and large-scale
silicene.
\end{abstract}

\maketitle

\textbf{Introduction} With the development of semiconductor
industry toward smaller scale, the rich quantum phenomena in
low-dimensional systems may lead to new concepts and
ground-breaking applications. In the last decade graphene has
emerged as a low-dimensional system for both fundamental research
and novel applications including electronic devices, energy
storage and transparent protection layer
\cite{Geim,Neto,Pumera,Kim}. Inspired by the fruitful results
based on graphene, recently a lot of interest has been drawn to
group IV (Si, Ge) analogs of graphene.
\cite{Cahangirov,Kara1,Kara2} It has been theoretically shown that
silicene, with Si atoms packed in a honeycomb lattice like
graphene, is a new massless Dirac Fermion system.
\cite{Cahangirov,Liu1} Compared with that of graphene, the
stronger spin-orbit coupling in silicene may lead to detectable
quantum spin Hall effect (QSHE) and other attractive properties.
\cite{Kane,Bernevig,Liu1,Liu2} The compatibility of silicene with
silicon-based nanotechnology makes this material particularly
interesting for device applications.

As the theoretical studies on silicene are rapidly increasing, the
major challenge in this field is now the preparation of high
quality silicene films. However, to date, there is still no solid
evidence for the observation of a silicene film. There have been a
few works on the formation of silicene nanoribbons on Ag(110) with
graphene-like electronic signature \cite{Aufray,Padova}. The only
published work on the preparation of silicene-like sheets was
reported by Lalmi et al. on Ag(111) \cite{Lalmi}. They showed
scanning tunneling microscopy (STM) images of honeycomb monolayer
structure that resembles monolayer graphene structure. However, in
their experiment the observed lattice constant was about 17\%
smaller than the theoretically proposed model or the value of bulk
silicon. Such a huge compression of the lattice is rather unlikely
to be induced by the strain between the film and the substrate.
Their results therefore remains to be confirmed and understood.
For the purpose of finding evidence of silicene, and optimizing
the preparation procedure for growing high quality silicene films,
it is important to build a complete understanding of the formation
mechanism and growth dynamics of possible silicon structures on
Ag(111), which is currently believed to be the best substrate for
growing silicene.

In this paper, we present a systematic study of the self-organized
superstructures formed by sub-monolayer silicon grown on Ag(111),
by STM and scanning tunneling spectroscopy (STS). We found that,
depending on the substrate temperature and silicon coverage,
several monolayer superstructures can form on Ag(111). These
superstructures are distinct from any known surface structures of
bulk silicon, and are characterized by honeycomb building blocks
and structures. At sufficiently high temperature and Si coverage,
monolayer and multilayer silicene films were grown. Combined with
first-principles calculations, the structural models of these
phases are proposed and their evolution with temperature and Si
coverage is discussed. Our work provides a complete understanding
of the structure evolution of Si on Ag(111), which is desirable
for fabrication of high quality silicene and exploring its novel
physics and applications.

\textbf{Experiments and Methods} Experiments were carried out in a
home-built low temperature STM with a base pressure of
5$\times$10$^{-11}$ Torr. Single crystal Ag(111) sample was
cleaned by cycles of argon ion sputtering and annealing. Silicon
was evaporated from a heated wafer ($\approx$1200 K) onto the
pre-heated substrate. The deposition rate of silicon was kept at
0.08-0.1 ML/min (here 1 monolayer refers to the atomic density of
a ideal silicene sheet). The STS data were acquired using a
lock-in amplifier by applying a small sinusoidal modulation to the
tip bias voltage (typically 10 mV at 676 Hz). All our STM
experiments were carried out at 77 K.

First-principles calculations were performed within the framework
of density functional theory (DFT) using Projected Augmented Wave
(PAW) \cite{Vanderbilt,Blochl} pseudopotentials and the
Perdew-Burke-Ernzerholf (PBE) \cite{Perdew} form for
exchange-correlation functional, as implemented in Vienna ab
initio simulation package (VASP) \cite{Kresse}. During
calculations, the structures were relaxed without any symmetry
constraints using a plane-wave energy cutoff of 250 eV. The
convergence of energy is set to 1.0$\times$10$^{-4}$ eV. The
relaxation process continues until forces are below 0.01 eV/\AA.

\textbf{Results and Discussions} Silicon atoms deposited on
Ag(111) tend to form clusters or other disordered structures when
the substrate temperature is below 400 K during growth (data not
shown here). As substrate temperature increases to 420 K, two
ordered phases form, as shown in Fig. 1. The less ordered phase
consists of close packed protrusions (labeled T), and the highly
ordered phase exhibits honeycomb structure (labeled H). The two
phases can coexist on the surface within a large coverage range,
from 0.5 ML [Fig. 1(a)] to 0.9 ML [Fig. 1(b)].

\begin{figure}[tbp]
\includegraphics[width=8.5cm,angle=-0]{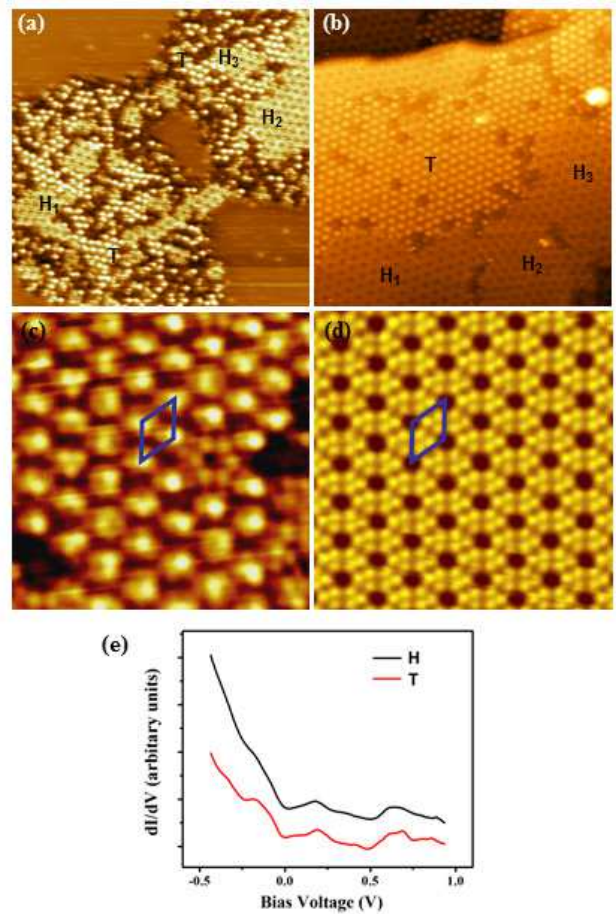}
\caption{(color online) (a) STM image (V$_{tip}$ = 1.2V) of 0.5 ML
silicon atoms deposited on Ag(111) surface at substrate
temperature of 420 K. The areas with phase T were marked by "T",
while the areas with phase H (with different rotation angles) were
marked by H$_1$-H$_3$, respectively. (b) STM image (V$_{tip}$ =
-1.5V) of 0.9 ML silicon atoms deposited on Ag(111) surface at
substrate temperature of 440 K. The areas of phase T and H are
labelled. Notably H$_1$-H$_3$ mark areas with phase H in different
orientations. (c) and (d) High resolution STM images
(8.5$\times$8.5 nm$^2$, V$_{tip}$ = -1.0V) showing the atomic
structure of phase T and H, respectively. The blue rhombuses in
(c) and (d) indicate the unit cells of two phases. (e) dI/dV
spectra taken at areas of phase T (red) and H (black)
respectively. The curves have been shifted vertically for
clarity.}
\end{figure}

The coexistence of phase H and T indicates that these two phases
have quite similar formation energy and stability. One can
therefore expect similarity and relations between the atomic
structure of these two phases. Indeed, the high resolution STM
images in Fig. 1(c) and (d) show that every big bright protrusion
in both phase T and H is indeed composed of three smaller spots
that we refer to as a "trimer", although in phase H the trimers
are perfectly ordered while in phase T they exhibit some
irregularity and distortion when looked closely. The two phases
share the same periodicity of 1.18 nm, therefore the density of
trimers in phase H is just twice of that in phase T. We further
performed scanning tunneling spectroscopy (STS) measurements for
the two phases [Fig. 1(e)]. The dI/dV curves exhibit very similar
features of electronic density of states (DOS), which strongly
implies that the two phases may share some common building blocks
in their atomic structures. In fact, in Fig. 1(c), there is a
noticeable point showing a corner hole and six protrusions
surrounding it -- a characteristic signature of formation of phase
H. Moreover, we notice that phase T prefers to form at lower Si
coverage, and slightly lower temperature as compared with phase H.
With the increase of substrate temperature and coverage, phase T
decreases in percentage and eventually disappear completely at 460
K. Meanwhile, phase H can spread over the surface if the coverage
is sufficiently high. This means that phase H is more stable than
phase T, and phase T can be regarded as a precursor state of phase
H.

We now face a direct question whether these two phases, especially
the well-ordered H phase, are the theoretically proposed silicene.
The phase T can be excluded first due to the significantly lower
density of Si than that of phase H. We notice that the periodicity
of 1.18 nm, is almost exactly four times the lattice constant of
Ag(111)-1$\times$1 surface, 0.29 nm, or three times the lattice
constant of silicene, 0.38 nm. Therefore both phases H and T can
be written as 4$\times$4 reconstruction with respect to the
1$\times$1-Ag substrate, or 3$\times$3 reconstruction with respect
to 1$\times$1 silicene lattice (in this paper we refer to as
3$\times$3). In fact, Ag/Si system is known as a typical "magic
mismatched" system, such that three times the lattice constant of
Si equals exactly four times the lattice constant of Ag. If one
assume the observed H phase to be the theoretically proposed
silicene, it is possible to obtain a 3$\times$3 superstructure by
placing the silicene lattice in parallel with the 1$\times$1-Ag
lattice. However, the crucial point in such moir\'e pattern models
is that: the periodicity of the superstructure, or essentially the
moir\'e pattern, is strictly linked with the relative orientations
of the two overlapping lattices. If one obtain a 3$\times$3
moir\'e pattern in one orientation, it will be impossible to
observe the same pattern in another inequivalent crystallographic
orientation. However, as we show in Fig.1(a), we have observed the
formation of 3$\times$3 domains on the same Ag terrace, with
different orientations which are obviously inequivalent. This
simple experimental fact excludes the possibility that the
3$\times$3 structure is a silicene lattice placed on
1$\times$1-Ag. In another word, the 3$\times$3 reconstruction
should come from the structure of the overlayer itself, instead of
from the commensuration between the overlayer and the substrate.
It is, however, noticeable that the 3$\times$3 reconstruction is
most clearly resolved in one major crystallographic orientation,
while in other orientations some irregular distortion of the
lattice is seen, which should come from the influence of substrate
Ag lattice.

As noted above, the periodicity of 1.18 nm is three times the
lattice constant of Si(111), 0.38 nm, which is also close to the
calculated lattice constant of silicene. \cite{Liu1} Based on STM
observation of the characteristic corner hole structure, we
propose a model of phase H as shown in Fig. 2(a). In this model,
the corner holes are due to missing of a hexagonal silicon rings
in each 3$\times$3 cell of a complete honeycomb silicene
structure. This model has been confirmed by first-principles
calculations. In the calculation the structure was modelled with
low-buckled silicene lattice \cite{Liu1} with missing silicon
rings at the corners. The six Si atoms around the corner are
hydrogenated. After relaxation, there are no in-plane changes of
the position of silicon atoms, but the atoms close to the corner
holes (red atoms in Fig. 2(a)) move upward, corresponding well
with the trimer feature observed by STM.

\begin{figure}[tbp]
\includegraphics[width=8.5cm,angle=-0]{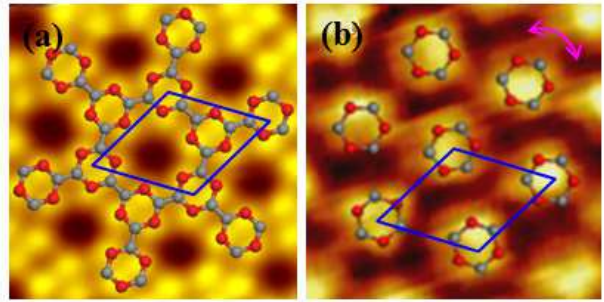}
\caption{(color online) (a) and (b) High resolution STM images
superposed with calculated model of phase H and T, respectively.
The red and grey balls in the models represent buckled and
unbuckled silicon atoms, respectively. The blue rhombus in (a) and
(b)indicate unit cells as shown in Fig. 1(c) and (d). The double
arrows indicate that the rings can rotate randomly along their
centers.}
\end{figure}

Based on the atomic structure of phase H, the understanding of the
atomic structure of phase T becomes straightforward. Because the
STM observation shows that the density of Si trimers in phase T is
half of that in phase H, we construct the model of phase T by
removing half of the silicon rings in phase H, leaving only one
hexagonal silicon ring per 3$\times$3 unit cell, as shown in Fig.
2(b). This model has also been validated by first-principles
calculations. Similar to the calculation of phase H, we chose
low-buckled silicene rings, saturated by hydrogen atoms as the
original structure. The calculation results show that this model
is stable. Each trimer corresponds to a buckled silicon ring with
three Si atoms moving upward. Such a structure can be considered
as self-assembly of hexagonal silicon rings stabilized by weak van
der Waals force. Compared with the honeycomb arrangement of Si
rings connected by covalent bonds in phase H, the weak connection
of Si rings in phase T might explain the observed more disordered
trimer structure, as compared with the highly ordered trimer
structure in phase H.

When the substrate temperature during silicon growth reaches 480
K, the silicon structure exhibits another phase with obvious
moir\'e pattern, which is long range ordered and can spread over
the whole surface as shown in Fig. 3(a). The orientation of
moir\'e pattern is along the $\langle1\bar{1}0\rangle$ direction
of Ag(111), and the period is about 3.8 nm. The high resolution
STM image in Fig. 3 indicates that a few complete honeycomb rings
with lattice period about 1.0 nm is observed at the bright part of
the moir\'e pattern, and the other parts are rather defective and
disordered. Additionally the angle between the direction of
moir\'e pattern and honeycomb structure is about $30^{\circ}$. The
dI/dV spectra measured on this structure shows a peak at 0.3V and
a shoulder at 0.9V, which is distinct from that of phase T and H,
and indicating an essentially different structure formed.

Considering that the complete honeycomb structure with 1.0 nm
periodicity is only observed at special positions on surface, we
proposed that this honeycomb superstructure consists of fragments
of single layer of silicene with strong interaction with the
Ag(111) substrate. In order to clarity our supposition,
first-principle calculations have been performed. The structure
model is constructed as single layer, low-buckled silicene being
in registry with five Ag(111) planes. Except for the two bottom Ag
layers, all atoms are relaxed during the geometry optimization.
The energetically stable structure is shown in Fig. 3(d). From the
calculation results we find that silicon atoms directly above a
silver atom (red balls in Fig. 3(d)) are higher than other silicon
atoms. As a result these atoms should be observed as bright
protrusions in STM image and forming a $\sqrt{7}\times\sqrt{7}$
superstructure with respect to silicene or
(2$\sqrt{3}$$\times$2$\sqrt{3}$)R$30^{\circ}$ superstructure with
respect to Ag(111). This gives a larger honeycomb lattice with
period 0.386$\times\sqrt{7}$ = 1.02 nm, in accordance well with
our experimental data. The simulated STM image according to the
calculated model is shown in Fig. 3(f). The similar structure
features and lattice period as observed in experimental STM images
(Fig. 3(e)) strongly support our suggested model. Actually there
is a slight deviation between the lattice constant of
$\sqrt{7}\times \sqrt{7}$ (1.02 nm) superstructure of silicene
from that of 2$\sqrt{3}$$\times$2$\sqrt{3}$ (1.00 nm) lattice of
Ag(111), which result in the formation of the moir\'e pattern. The
optimized structural model in Fig. 3(d) shows the hexagonal rings
of silicene are twisted due to the strong interaction between
silicon atoms and silver substrate. The bright parts of moir\'e
pattern is where the positions of atoms in silicene are little
deviated from that of Ag(111), which make the honeycomb
superstructure stable enough to keep the hexagonal rings complete.
In other parts of moir\'e pattern, the larger deviation of
position between atoms in silicene and those of Ag(111) lead to
unstable honeycomb structure and eventually breaks the hexagonal
rings of silicene, resulting in defective and disordered
structures. The disordered structures were not obtained in our
calculations because the unit cell we choose is much smaller than
that of a moir\'e pattern. The angle between the lattice direction
of the $\sqrt{7}\times\sqrt{7}$ superstructure and
$\langle1\bar{1}0\rangle$ direction of Ag(111) is $30^{\circ}$, so
the angle between the direction of moir\'e pattern and
$\langle1\bar{1}0\rangle$ direction of Ag(111) should be zero,
which has been confirmed by our experiments.

\begin{figure}[tbp]
\includegraphics[width=8.5cm,angle=-0]{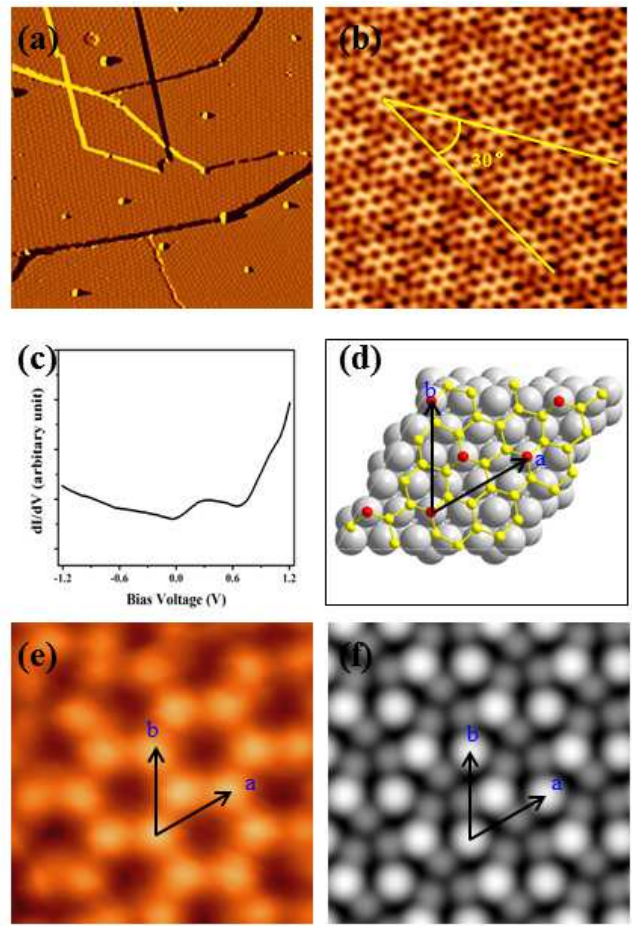}
\caption{(Color online)(a) A derivative STM image (200$\times$200
nm$^2$, V$_{tip}$ = 1.43V) of 0.9 ML silicon atoms deposited on
Ag(111) surface at substrate temperature of 480K. (b) High
resolution STM image (15$\times$15 nm$^2$, V$_{tip}$ = -1.0V)
showing the atomic structure of moir\'e patterns. The bright areas
exhibit complete honeycomb rings with period of 1.0 nm while other
areas are defective and disordered. The angle between the
orientation of the hexagonal rings and the direction of moir\'e
patterns is $30^{\circ}$. (c) dI/dV spectra taken at the moir\'e
pattern phase, in which a peak at 0.3V and a shoulder at 0.9V are
observed. (d) Calculated model of $\sqrt{7}\times\sqrt{7}$
superstructure of silicene. The grey, yellow and red balls
represent the silver atoms, lower silicon atoms and higher silicon
atoms, respectively. (e) and (f) Experimental and simulated STM
images (1.0 eV above Fermi energy) showing the similar structure
features and unit cell of lattice.}
\end{figure}

As the substrate temperature reaches 500 K and the coverage is up
to 0.8 ML, we observed dense honeycomb structure which we identify
as silicene. The STM image in Fig. 4(a) shows a one-atom-thick
silicene sheet across the step edges of the Ag(111) surface
without losing continuity of the atomic lattice, which is similar
to graphene grown on metal surfaces \cite{Coraux}. The high
resolution STM image of Fig. 4(b) shows a honeycomb structure.
However, different from the reported 1$\times$1 structure of
silicene \cite{Lalmi}, the lattice period of the honeycomb
structure we observed is about 0.64 nm, which is corresponding to
a $\sqrt{3}\times\sqrt{3}$ honeycomb superstructure with respect
to the 1$\times$1 silicene lattice. This superstructure can be
explained by a symmetric-buckled silicene model \cite{Chen} which
is shown in Fig. 4(c). The calculation of free standing silicene
shows that the six silicon atoms in one hexagonal ring are not in
plain: two atoms are buckled upward and one atom is buckled
downward, forming AB\=A configurations. The upper buckled atoms
are resolved by STM as the $\sqrt{3}\times\sqrt{3}$ honeycomb
superstructure. Different from graphene epitaxially grown on
metals \cite{Land,Marchini}, we did not observe moir\'e patterns
in our film, which may originate from the weak interaction between
silicene and the metal substrate \cite{Sutter1}. A typical dI/dV
spectrum obtained at the silicene terrace (black curve in Fig.
4(e)) shows a shoulder at 0.3V and a peak at 0.9V, which is
similar as the LDOS distribution measured on moir\'e patterns
phase. Another remarkable feature is a small dip located at 0.5 eV
which is corresponding to the Dirac point (DP) of silicene. The
dip is not much obvious compared with that of graphene
\cite{Zhang,Zhao}, which is probably due to the pronounced
electronic DOS of the underlying Ag(111) substrate superimposed on
the dI/dV spectra. The deviation of the energy position of DP from
the Fermi energy may stem from the charge transfer from the
Ag(111) surface to silicene.

\begin{figure}[tbp]
\includegraphics[width=8.5cm,angle=-0]{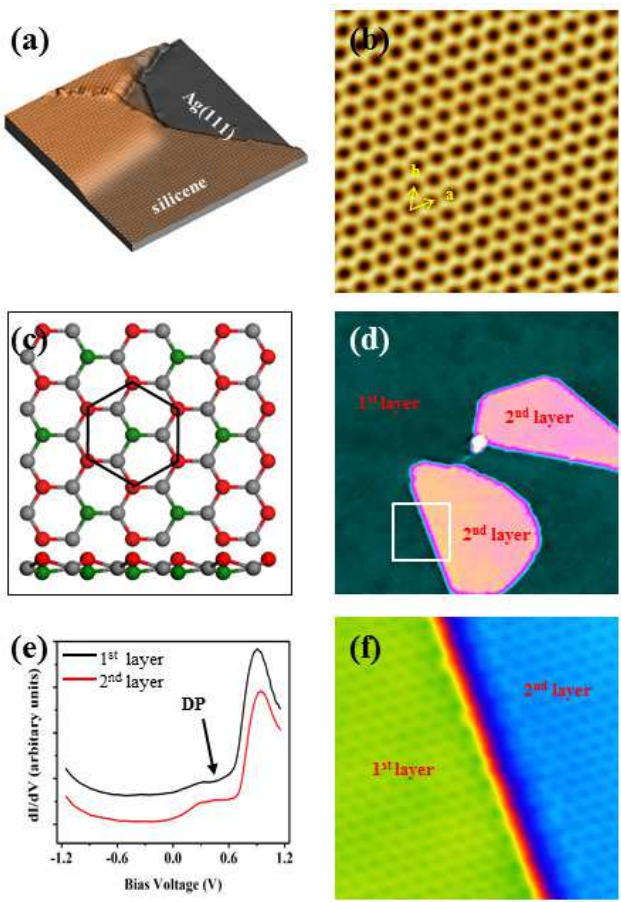}
\caption{(Color online) (a) 3D STM image (30$\times$30 nm$^2$,
V$_{tip}$ = 1.0V) of a single layer of silicene island across a
step edge of Ag(111). (b) High resolution STM image (8$\times$8
nm$^2$, V$_{tip}$ = 1.2V) of one monolayer silicene terrace
showing the $\sqrt{3}\times\sqrt{3}$ honeycomb superstructure with
the period of 0.64 nm. (c) Top and side view of schematic model of
$\sqrt{3}\times\sqrt{3}$ superstructure of silicene. The red, grey
and green balls represent the upper buckled, in-plain, and lower
buckled Si atoms, respectively. The $\sqrt{3}\times\sqrt{3}$
honeycomb superstructure is indicated by the black hexagon. (d)
STM image (54$\times$54 nm$^2$, V$_{tip}$ = 1.5V) of 1.2 ML
silicon atoms deposited on Ag(111) surface at substrate
temperature of 500K showing second layer of silicene formed on the
first layer of silicene. (e) dI/dV spectra taken on the first
(black) and second (red) layer of silicene respectively. (f) High
resolution STM image (10.5$\times$10.5 nm$^2$, V$_{tip}$ = 1.5V)
of area as marked by the white rectangle in (d) showing atomic
structure of the first and second layer of silicene
simultaneously.}
\end{figure}

The assignment of the above phase as silicon gets a direct proof
by the observation of second layer of silicene at higher coverage,
as shown in Fig. 4(d). The high resolution STM image of Fig. 4(f)
shows the atomic structure of the first layer and second layer
silicene simultaneously. It is obviously that the second layer
silicene also exhibit a $\sqrt{3}\times\sqrt{3}$ honeycomb
superstructure, which indicates that the $\sqrt{3}\times\sqrt{3}$
honeycomb superstructure should originate from free standing
silicene and not influenced by Ag(111) surface. This layer-stacked
silicon structure is similar as graphite, and is a new structural
phase of silicon, which may host many novel properties. The dI/dV
spectrum on second layer of silicene (red curve in Fig. 4 (e))
resembles that on first layer. This striking similarity between
the LDOS of monolayer and bilayer silicene can also confirm our
predation that interactions between the monolayer silicene and
Ag(111) are as weak as that between the two silicene layers.

Even at substrate temperature about 500 K during growth, silicon
atoms tend to form the moir\'e pattern phase if the coverage of
silicon is considerably less than 0.8 ML. This indicates that the
atomic density of silicene is higher than that of the moir\'e
pattern phase, which justifies our structural model again.
Increasing the substrate temperature to above 600 K, no structure
of silicon can be observed anymore, leaving only a bare Ag(111)
surface. Furthermore, if the sample of silicene on Ag(111) is
annealed up to 600 K, silicene film will also disappear. The upper
temperature limit that our silicene film can endure is
considerably lower that that of graphene \cite{Land,Sutter2,Yu}.
This is another evidence of the weak interaction between silicene
and Ag(111) substrate.

\textbf{Conclusion} We have systematically investigated the
structure evolution of silicene on Ag(111). With the increase of
the substrate temperature, silicon atoms on Ag(111) overcome
potential barriers and form some metastable structural phases such
as self-assembled honeycomb building blocks (phase T and H) and
incomplete silicene film (moir\'e pattern structure). The most
stable phase, decoupled monolayer and bilayer silicene film, will
appear eventually. This work provides methods to fabricate high
quality silicene, which is essential to investigate its novel
properties, and brings it closer to the use in nanotechnology and
other related areas.

\textit{Acknowledgements}: This work was supported by the NSF of
China (Grants No. 11074289, 91121003), and the MOST of China
(Grants No. 2012CB921703, 2009CB929101)..

\end{document}